\documentclass[5p, sort&compress]{elsarticle}
\usepackage{amsmath}
\usepackage[american]{babel}
\usepackage[T1]{fontenc}
\usepackage[utf8x]{inputenc}
\usepackage{graphicx}
\usepackage{hyperref}
\usepackage{comment}
\usepackage{footnote}
\usepackage{tikz}

\newcommand{\mytitle}{Interface-controlled creep in metallic glass composites}

\hypersetup{
  final,
  pdfborder={0 0 0},
  pdftitle={\mytitle},
  pdfauthor={C. Kalcher, T. Brink, J. Rohrer, A. Stukowski, K. Albe},
  colorlinks=true,
  urlcolor=blue,
  linkcolor=blue,
  citecolor=blue
}

\begin{document}

\title{\mytitle}

\author[tu]{Constanze Kalcher\corref{cor1}}
\ead{kalcher@mm.tu-darmstadt.de}
\address[tu]{Institute of Materials Science, Technische Universit{\"a}t
Darmstadt, Jovanka-Bontschits-Str. 2, D-64287 Darmstadt, Germany}

\author[tu]{Tobias Brink}
\author[tu]{Jochen Rohrer}
\author[tu]{Alexander Stukowski}
\author[tu]{Karsten Albe}
\cortext[cor1]{Corresponding author}

\date{\today}%

\begin{abstract}
  In this work we present molecular dynamics simulations on the creep behavior of $\rm Cu_{64}Zr_{36}$ metallic glass composites. 
  Surprisingly, all composites exhibit much higher creep rates than the homogeneous glass. The glass--crystal interface can be viewed 
  as a weak interphase, where the activation barrier of shear transformation zones is lower than in the surrounding glass.
  We observe that the creep behavior of the composites does not only depend on the interface area but also on the orientation 
  of the interface with respect to the loading axis. We propose an explanation in terms of different mean Schmid factors of the 
  interfaces, with the amorphous interface regions acting as preferential slip sites. 
\end{abstract}


\journal{Acta Materialia, online at
  \href{https://doi.org/10.1016/j.actamat.2017.08.058}
       {\texttt{https:/\!/doi.org/10.1016/j.actamat.2017.08.058}}.}

\maketitle
\begin{tikzpicture}[remember picture,overlay]
  \node [anchor=north west, font=\footnotesize\itshape, align=left,
         xshift=-\oddsidemargin, yshift=1in]
        at (current page.south west)
        {\phantom{Pp}\\
         © 2018. This manuscript version is made available under the
         CC-BY-NC-ND 4.0 licence.\\
         \textup{\url{http://creativecommons.org/licenses/by-nc-nd/4.0/}}};
\end{tikzpicture}%
\vspace{-2.5\baselineskip}
\section{Introduction}
  \begin{figure*}[t]
    \centering
    \includegraphics[width=0.85\textwidth]{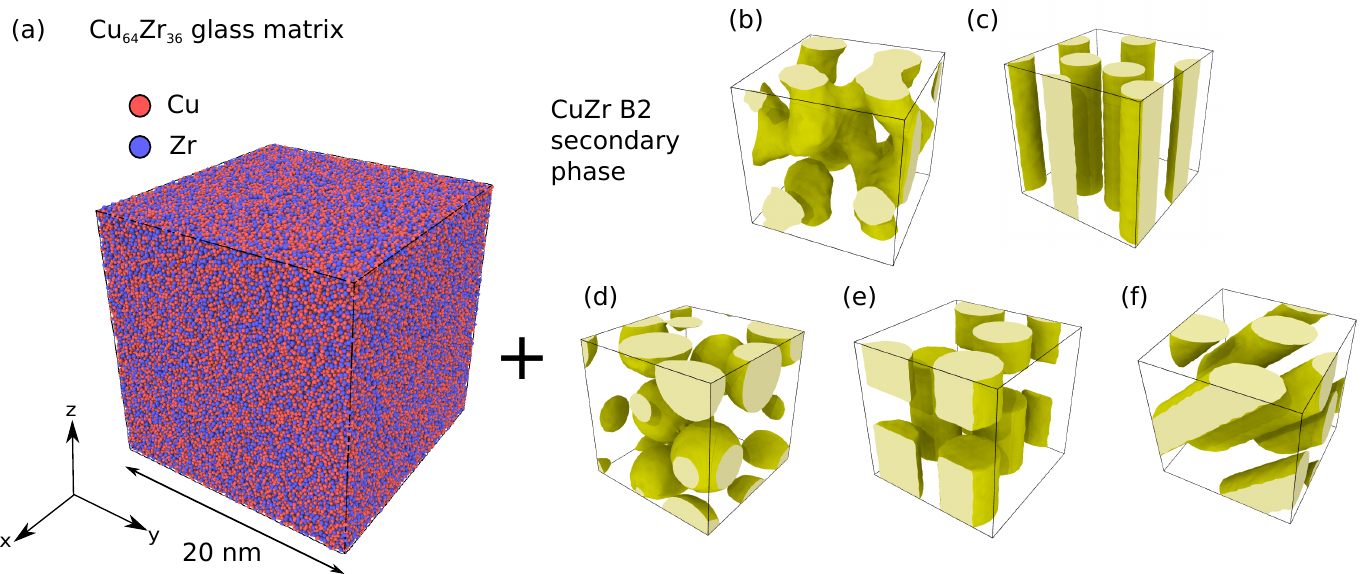}	
    \caption{Snapshots of the composites from the sample series with  $f=30\, \%$ and $\phi = 210 \, \rm \mu m^{-1}$. (a) The glass sample. Color coding:
      Blue and red represent Cu and Zr atoms, respectively. (b)--(f) Surface meshes of the secondary CuZr B2 phases that are inserted into the glass matrix.}
    \label{fig:f1}
  \end{figure*}
 
  Creep in homogeneous metallic glasses (MGs)~\cite{Inoue2000, Wang2004} can be understood in the framework of shear transformation zones
  (STZs)~\cite{Argon1979, Schuh2007}. Classically, three different deformation regimes in
  terms of stress, temperature and strain rate are distinguished~\cite{Schuh2007, Hufnagel2016}: elastic deformation, homogeneous
  plastic deformation at high temperatures and low stresses, and heterogeneous plastic deformation (shear banding) at low temperatures
  and high stresses.  All plastic deformation is carried by STZs~\cite{Argon1979, Falk1998, Albano2005, Schuh2007}, only their
  localization is different.  Moreover, homogeneous deformation at high temperatures can be understood as a creep-like phenomenon~\cite{Heggen2005}.
  In MGs it usually sets in at about $0.6$--$0.7 \ T_g$~\cite{Schuh2007}, which for some alloys is not that 
  far above room temperature. Indeed, it was shown that creep-like homogeneous deformation in Cu-Zr MGs occurs even at room temperature and 
  stresses close to yield. This regime is often designated as elastic~\cite{Park2008, Lee2008}, yet, the extent 
  of this elastic regime is unclear, since a continuum of activation barriers (starting from very low values) was found in 
  glasses~\cite{Rodney2009, Derlet2011,  Derlet2016}. This suggests that the ``elastic regime'' may be subject to slow, creep-like 
  deformation, making the difference between elastic and homogeneous flow a matter of timescale.

  In nanoindentation~\cite{Falk1998, Yoo2010, Yoo2010a, Li2008, Castellero2008, Huang2009} and compression tests~\cite{Heggen2005}
  on different glass alloy systems, homogeneous creep deformation has been observed and stress exponents have been measured. 
  However, Li \textit{et al}.~have pointed out the unreliability of the stress exponent extracted from nanoindentation data
  of MGs~\cite{Li2016}. Song \textit{et al}.~performed high temperature compression tests on Au-based MG micro-pillars~\cite{Song2009} 
  and comparison of their results with Argon's model for creep in metallic glasses allowed the extraction of activation energies. These results at least 
  support the view that creep in metallic glasses resembles the homogeneous deformation as postulated by Argon~\cite{Argon1979, Argon1983}.  Molecular dynamics (MD) simulations of a 2D and 3D binary glass support this mechanistic picture of homogeneous 
  deformation~\cite{Falk1998, Albano2005}. Interestingly, for CuZrNiAl MGs it has been found that high-temperature mechanical creep
  has a positive effect on their room temperature plasticity~\cite{Dmowski2013}: While an annealing treatment at 
  $300 \, ^{ \circ} \rm C$ leads to brittleness, creep processing at the same temperature was even able to restore plasticity.

  For composites based on amorphous matrices, less data is
  available. In metallic systems, the focus of recent efforts
  has been the improvement of ductility in the low-temperature regime,
  where shear banding is predominant \cite{Hays2000, Hays2001,
    Lee2004, Lee2006, Eckert2007, Hajlaoui2007, Fu2007, Pauly2010,
    Albe2013, Brink2016}. While these composites are successful at
  modifying shear band propagation, few investigations on the role of the secondary
  crystalline phases in the higher temperature, creep-like regime exist.
  A 
  metallic glass composite of $\rm Zr_{55}Al_{10}Cu_{30}Ni_{5}$ containing nanocrystalline W particles has been synthesized, 
  and found to exhibit an increased viscosity with increasing W content~\cite{Eckert1999}. 
  In polymer science, partially crystalline samples are common and could be referred to as ``composites'', but
  they are mostly discussed in terms of viscoelasticity~\cite{Hutchinson1995}.  The creep of composites
  based on amorphous matrices has so far only been investigated for ceramic systems: Recently, it has been found 
  that silicon oxide glasses~\cite{Rouxel2001, Papendorf2012, Ionescu2014} exhibit a high refractoriness with respect to 
  creep which makes them potential candidates for high-temperature applications. Several structure models have been proposed 
  for these polymer derived ceramics -- picturing a silica matrix and a secondary phase of excess carbon,
  either in the form of spherical graphitic nanodomains  or a graphene-like network structure~\cite{Saha2006, Ionescu2014}.
  But overall, information on the underlying mechanism of creep in these systems is still sparse.  Analogous to
  crystalline materials, the question arises if mechanical reinforcement against creep is possible by introducing a stronger 
  non-yielding secondary phase, e.g., by combining a glass matrix with a crystalline phase. Furthermore, evidence for effects of the geometry of secondary phases were already found: The heat conductivity in a copper-fiber-reinforced ZrTiCuNiBe bulk metallic glass exhibits a significant anisotropy depending on the orienation of the fibers~\cite{Wadhwa2007}.

  Within this context  we want to shed 
  light on the question if a non-deformable crystalline secondary phase embedded in a MG can increase the
  creep resistance of the MG.  We are especially interested in the influence of the topology of the secondary phase.
  To this end, we need to be able to accurately treat the individual constituents of the composites, especially the interface. This
  goes beyond the possibilities of continuum methods, since it would require prior knowledge of the constitutive laws at the
  interface. Therefore we resort to atomistic molecular dynamics simulations, which only need a description 
  of interatomic interactions and a structure as an input. As a model material, we choose the binary glass former 
  Cu$_{64}$Zr$_{36}$, which is well described by EAM potentials~\cite{Mendelev2009}. In order to asses the
  impact of a secondary phase on the creep behavior, we study various geometries of an embedded crystalline phase.  
  In Cu--Zr glasses, B2 CuZr precipitates can be obtained by annealing procedures and during deformation
  ~\cite{Jiang2006, Jiang2007, Pauly2007, Pauly2009} and are thus chosen as a representative crystalline phase for our model systems.
  The B2 phase is brittle and dislocation activation is hindered by the high antiphase-boundary energy. 
  In previous simulations~\cite{Brink2016}, this phase was shown to only respond elastically when part of a composite.

  Therefore, our systems are a mix of a crystalline phase with high yield stress, that only reacts elastically,
  combined with an amorphous phase. As stated earlier, such systems were also proposed for ceramics~\cite{Papendorf2012}, 
  and are therefore relevant models even beyond metallic systems.

\section{Computational methods and analysis}

  We model the $\rm Cu_{64}Zr_{36}$ metallic glasses as well as the glass--crystal composites with a CuZr B2 phase using 
  the MD code \textsc{lammps}~\cite{Plimpton1995} and the \textsc{eam}-potential by Mendelev \textit{et al.}~\cite{Mendelev2009}. 
  The time step for the integration of the equation of motion is $2 \, \rm fs$ in all simulations.  Temperature and pressure 
  controls are realized by a Nos\'e--Hoover thermostat and a Parinello--Rahman barostat.  Periodic boundary conditions are applied
  in all simulations.

  Figure~\ref{fig:f1} shows a snapshot of the homogeneous $\rm Cu_{64}Zr_{36}$ metallic glass and the different secondary phase
  topologies used in this study.  The matrix glass, a cubic box of dimension $20 \times 20 \times 20 \, \rm nm^3$ of
  $\rm Cu_{64}Zr_{36}$, is obtained by quenching from the melt with $0.01 \, \rm K/ps$ at ambient pressure. This glass is then equilibrated
  at $300 \, \rm K$, again at ambient pressure, for $1 \, \rm ns$.
  The composites are created starting from the glass matrix, by cutting out the desired geometry for the secondary phase  and filling
  the empty space with a B2 crystal oriented with the $\langle 100 \rangle$ directions along the cartesian axes.  The lattice constant for B2 was
  obtained using a simulation at $300\,\mathrm{K}$.  After assembly, the composites are relaxed  for $1 \, \rm ns$  at $300 \, \rm K$
  and ambient pressure.

  We start with an interpenetrating, continuous network, depicted in Fig.~\ref{fig:f1}(b), which serves as a model for percolating
  crystalline structures embedded in a glass. Such structures resemble those that have been proposed, e.g., for phase-separating 
  oxide glasses~\cite{Papendorf2012}.  The networks in our simulations were obtained by creating a mask using simulated spinodal 
  decomposition of a binary alloy as described in Ref.~\cite{Ngo2015}. The two components are chosen with volume fractions of 
  $f_{A}=0.3$ and $f_{B}=0.7$. The system completely decomposes into a network (A) and a matrix phase (B). Then, the network phase
  starts to coarsen. The network structure is extracted from four different evolution stages of
  this coarsening process and used as a mask. Thus, the masks have the same volume fraction, $f = 0.3$, but different surface areas.  
  We use those two parameters to describe the geometry of the secondary phase: The crystalline volume fraction $f$ and the specific 
  interface area $\phi$. We define the latter as the surface area of the free-standing crystalline phase divided by the total volume 
  of the composite:
    \begin{equation}
      \phi  = \frac{A_\text{interface}}{V_\text{tot}} \; .
    \end{equation}
  The surface area is determined using a surface-mesh construction by Delaunay tesselation~\cite{Stuko2014} as implemented in the 
  software \textsc{ovito}~\cite{ovito}. For the networks, we choose four molds at different stages of the spinodal decomposition with
  specific interface areas $\phi_1 = 170 \, \rm \mu m^{-1}$, $\phi_2 = 210 \, \rm \mu m^{-1}$,  $\phi_3 = 320 \, \rm \mu m^{-1}$ and 
  $\phi_4 = 410 \, \rm \mu m^{-1}$.

  In order to understand the influence of geometry in more detail, we also employed simpler geometric shapes, such as continuous
  nanowires in various orientations (Figs.\ \ref{fig:f1}(c) and~\ref{fig:f1}(d)), non-percolating nanorods (Fig.~\ref{fig:f1}(e)), and spheres 
  (Fig.~\ref{fig:f1}(f)). These were prepared using the same volume fraction and specific interface areas as the network composites.

  For the creep simulation, we choose a temperature of $500\,\mathrm{K}$, which corresponds to about $0.6 \, T_g$.  The
  samples are heated to the target temperature with a heating rate of $0.1 \, \rm K/ps$.  Creep tests are performed by instantaneously
  applying a constant tensile stress in $z$ direction for $40\,\mathrm{ns}$ while keeping the other dimensions at ambient
  pressure.  The creep rate is obtained through a linear fit of the creep simulation data between $t_1 = 25\, \rm ns$ and
  $t_2 = 40 \, \rm ns$.  Furthermore, a load-cycling test is simulated at $500 \, \rm K$ by instantly unloading the samples for
  $20 \, \rm ns$ and subsequently reloading for $20 \, \rm ns$.

  The simulation data is analyzed and visualized with \textsc{ovito} \cite{ovito} using the atomic shear strain~\cite{Falk1998, Shimizu2007} as
  a measure for the localized deformation. In addition, 
we analyzed the structure of the glass matrix and glass-crystal interfaces using the
Voronoi tessellation method \cite{Voronoi1908,Voronoi1908a,Voronoi1909,Brostow1998} as implemented in
\textsc{ovito} \cite{ovito}. The resulting polyhedra
are characterized by the Voronoi index $\langle n_3, n_4, n_5, n_6, \ldots \rangle$, where $n_i$ denotes the number 
of $i$-edged faces of the polyhedron. The relative occurrence of these polyhedra characterizes the glass state: 
In Cu--Zr, the fraction of copper-centered icosahedra with index $\langle 0,0,12,0\rangle$ was found to be a
reliable indicator for relaxation and strength of the glass \cite{Lee2007, Cheng2008}.

\section{Plastic deformation of glass--crystal composites}

    \begin{figure}[h!]
	\centering   
	\includegraphics[width=0.5\textwidth]{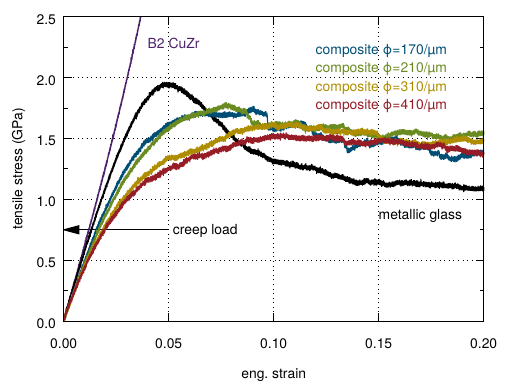}
	\caption{Tensile tests of the homogeneous glass, the single crystalline B2 CuZr, and the four composites with network
		  microstructure at $500 \, \rm K$. For all composites and the glass, a stress of $750 \, \rm MPa$ is significantly below 
		  the yield stress.}
    \label{fig:tensile}
    \end{figure}

  As a first step, we perform tensile tests on the different samples to ascertain that we choose a stress regime below yield for the 
  creep tests. Figure~\ref{fig:tensile} shows the stress-strain curves obtained by tensile tests of $\rm Cu_{64}Zr_{36}$ glass, 
  single-crystalline B2 CuZr, and the composites with a network structure [cf.~Fig.~\ref{fig:f1}(b)]. We perform 
  strain-rate-controlled tensile tests at $T=500\, \rm K$ with a constant engineering strain rate of 
  $\dot \epsilon  = 4\cdot 10^{7}/\rm s$. The tensile direction coincides with the $[001]$-orientation of the single-crystal.
  It can be seen that the Young's moduli of the glass and crystal are very similar, but the yield-stress of the glass is far 
  exceeded by the crystal's.
  Furthermore, it is evident that the four composites exhibit lower elastic moduli than their pure constituents, even though they 
  all contain the same volume fraction of crystalline network phase, $f = 30 \, \%$. This suggests that the specific interface area
  plays a strong role in the mechanical properties. The higher the specific interface area $\phi$, the lower is the composite's 
  mean elastic modulus. On the other hand, the composites have a higher flow stress and do not show the stress overshoot typical 
  for a glass. Such a ``flattening'' of the stress response has also been observed in other glasses with a microstructure containing 
  weakened interfaces, such as nanoglasses~\cite{Albe2013, Adjaoud2016}. From these results we can conclude that
  $\sigma = 750 \, \rm MPa$ is an appropriate stress level for creep testing at $ T = 500 \, \rm K$, since it is below the
  individual yield stresses of both the composite structures and the homogeneous glass, but high enough to observe creep on MD 
  timescales. The crystalline phase is still in the linear elastic regime up to strains of at least $5\, \%$.

\section{Creep simulations}

  \subsection{Homogeneous glass and interpenetrating network secondary phase}

    \begin{figure}[b!]
	\centering   
	\includegraphics[width=0.5\textwidth]{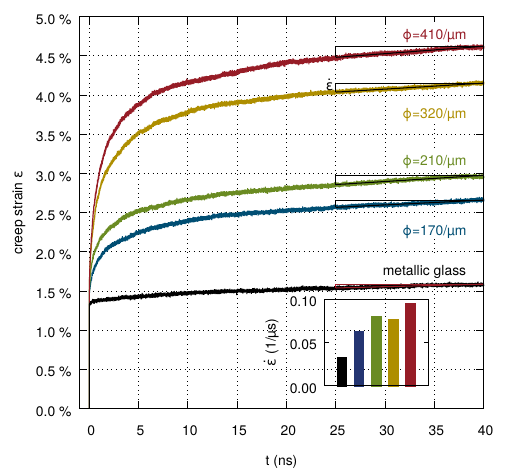}
	\caption{Dependence of true strain on loading time during the creep test of the composites with crystalline interpenetrating
		network microstructure. The creep curve of the  matrix glass is shown in black as a reference. The inset shows the
		different creep-rates of the samples fitted between $t=25 \, \rm ns$ and $t=40 \, \rm ns$.
		}
	\label{fig:creep-network}
    \end{figure}
    
  Figure~\ref{fig:creep-network} shows the creep curves of four network composites with topologies as depicted in Fig.~\ref{fig:f1}(b), with the parameters established above.
  Additionally, we show the creep curve of the homogeneous glass. At the simulation times available, we were never able to observe 
  failure. The creep curve rather suggests a mixed regime of creep stages~I and~II. Therefore, we simply extract the creep rate in all cases from the slope in the time range between  $t=25\, \rm ns$ and $t=40\, \rm ns$ to obtain comparable data. The creep strain depicted 
  in the following figures is always the true strain, $\epsilon_{\rm true}=\ln(L_z/L_{z0})$, where $L_z$ is the box length in 
  $z$ direction. In this way, we ensure that the fitted creep rates are independent of the initial box length $L_{z0}$ in 
  loading direction.

\begin{figure*}[t!]
      \centering
      \includegraphics[width=\textwidth]{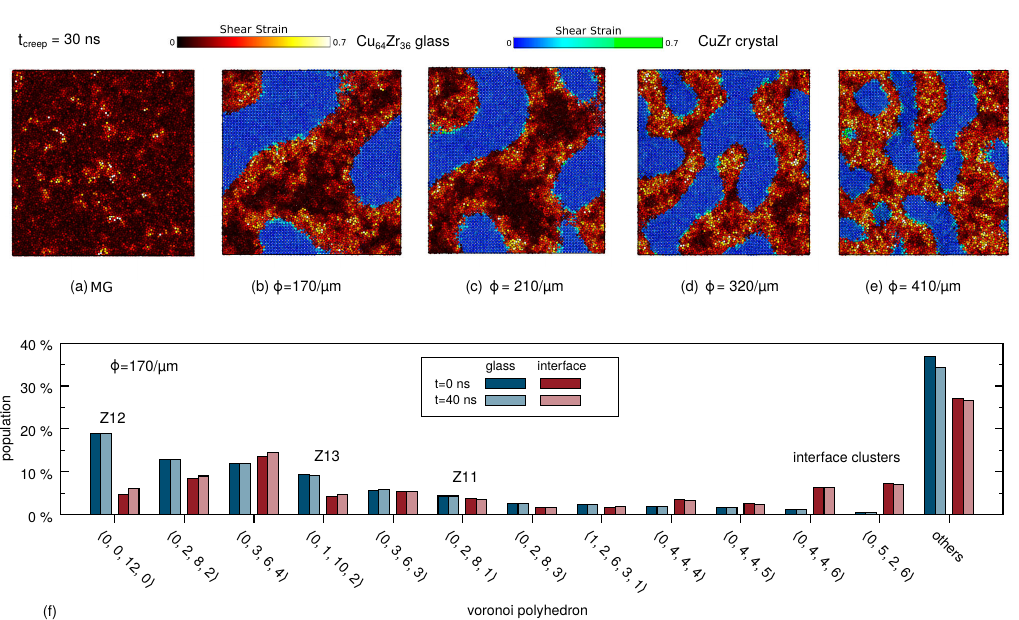}	
      \caption{Snapshots of (a) the MG  and (b)--(e) composites with secondary phase network ordered by increasing interface area  
	      after $t=30 \, \rm ns$ creep testing. The atoms are color coded according to the atomic shear strain, using two different color maps 
	      for the glass and the crystalline phase. Videos of these simulations are available in the supplemental material. 
	      (f) Voronoi statistics of the Cu atoms in the interface and glass matrix. Only Voronoi polyhedra that occur more
	      often than $2 \%$  are shown, the rest are summarized as ``others''. The energetically favorable Cu clusters with an
	      eleven-fold, twelve-fold and thirteen-fold coordination are indicated as Z11, Z12 and Z13 \cite{Ma2015}. 
	      The 12-fold signature motif $\left < 0,0,12,0 \right>$ and its defective variant $\left < 0,2,8,2 \right>$ are the most
	      prominent clusters in the glass matrix, but are severly reduced in occurence in the interface. The $\langle 0,4,4,6 \rangle$ and $\langle 0,5,2,6 \rangle$ clusters appear predominantly at the immediate glass--crystal interface.
      }
      \label{fig:snapshots}
    \end{figure*}

  Starting with the homogeneous glass, we find that it exhibits a comparatively slow creep rate: After an instant increase of $1.4\, \%$ elastic strain,
  the strain increases by less than $0.2 \, \%$ after $40 \, \rm ns$. The creep behavior of metallic glasses is well described by 
  Argon~\cite{Argon1979, Argon1983}. Assuming high temperatures and unpercolating STZs, the Argon model describes how the strain rate
  depends on stress $\tau$ and temperature $T$ in glasses:
    \begin{equation}
    \dot \epsilon(\tau,T)  = \alpha \nu_0 \gamma_0 \cdot \exp \left( -\frac{Q}{kT} \right)
    \sinh \left( \tau  \cdot \frac{ \gamma_0 \Omega}{ kT} \right) \; .
    \label{eq:argon}
    \end{equation}
  Here, $Q$ is the activation energy for plastic deformation, $\Omega$ is the volume of an STZ, $\gamma_0$ the characteristic shear
  strain of an STZ, $\nu$ its attempt frequency to flow and $\alpha$ a prefactor close to unity~\cite{Argon1980}.
  The local von Mises strain after $30 \, \rm ns$ of creep deformation is shown in Figure~\ref{fig:snapshots}(a). 
  The deformation is localized in clusters of a few atoms distributed homogeneously within the glass, 
  which is consistant with the STZ picture. This suggests that, at high temperatures and low stresses,
  the mechanism behind the deformation remains the same as for shear band formation, i.e., nucleation of 
  STZs. This is also corroborated by findings from Kassner \textit{et al.}~\cite{Kassner2015}, who summarized the activation 
  energies for creep of several metallic glasses. The values they found agree well with the activation energy of an STZ estimated 
  by Schuh \textit{et al.}~\cite{Schuh2007}, i.e.,\ $1$--$5 \rm \, eV$. For a $\rm Cu_{60}Zr_{40}$ glass, with a composition close 
  to the one used in this work, Argon and Kuo~\cite{Argon1980} measured an activation energy of about $2.4 \, \rm eV$.

    \begin{figure}[b!]
	\centering   
	\includegraphics[width=0.5\textwidth]{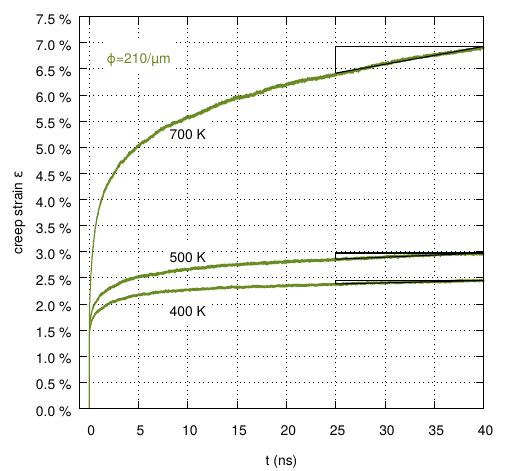}
	\caption{ Dependence of the creep behavior on temperature for the composite with a network structure such as in 
		  Fig.~\ref{fig:f1}.b with $\phi = 210 \, \rm \mu m^{-1}$ . The creep curves have been obtained for three 
		  different temperatures, $400\, \mathrm{K}$, $500\, \mathrm{K}$ and $700\, \mathrm{K}$. Both the total 
		  creep strain and creep rate strongly increase with temperature.
		  }
	\label{fig:400-700K}
    \end{figure}
  Moving on to the composites, we observe that the creep behavior's temperature dependence also resembles the behavior described 
  by Argon, cf.~Fig.~\ref{fig:400-700K}. Surprisingly,  Fig.~\ref{fig:creep-network} reveals that 
  the composites exhibit much higher creep rates than the homogeneous glass at the same temperature.  Furthermore, we note that the larger the
  specific interface area $\phi$, the higher is the total true creep strain after $40 \, \rm ns$, which is also reflected in the 
  inset to Fig.~\ref{fig:creep-network}, which shows that the creep rate increases with $\phi$. 
  For these glass--crystal composites, 
  composed of viscous glass phase and a stronger crystalline phase that cannot exhibit dislocation activity under the present testing 
  conditions, we would have expected a decrease in the creep rate. The observed increase seems counter-intuitive, considering
  the mix of a time-dependent deforming phase with a significant volume fraction of non-deformable phase. In fact,
  we showed above that at the creep strain levels obtained here in the composites, the crystalline phase only deforms elastically.
  Thus the creep behavior of the composite must be governed by the glass phase.

  How is it possible that introducing the network phase, makes the glass more prone to creep? A first explanation can be found 
  in the elastic moduli of the glass and crystal. While they do not differ much, the composites are less stiff than the homogeneous glass (see Fig.~\ref{fig:tensile}). Thus, we can infer that the composite interfaces
  must be weaker than the surrounding matrix. To understand the influence of the interface, we need to study the  creep deformation
  in the glass and crystal phase at the atomic level. Figures~\ref{fig:snapshots}(a)-(e) shows snapshots of the atomic strain in the
  homogeneous glass and composite structures. The color coding indicates the amount of atomic strain experienced by the atoms. 
  We use two different color maps for the atoms that initially belonged to the glass matrix or the B2 crystal. The results of
  this strain analysis support the hypothesis that the crystalline phase only deforms elastically since we could not observe 
  any dislocation activity. 
  Furthermore, Figs.~\ref{fig:snapshots}(b)-(e) reveal that the deformation is mostly localized in the interface. For very high specific
  interface areas ($\phi = 410\,\mathrm{\mu m}^{-1}$), almost the complete glass matrix can be considered as an interface region  
  because of the small distance between the two interfaces. The pure glass, however, exhibits a homogeneous distribution of STZs
  without the formation of shear bands, which resembles the non-interface glass regions in the composites.
  This suggests that the glass--crystal interface can be viewed as a weak interface interphase, where the effective activation 
  barrier for creep is lower than in the surrounding glass. Therefore, STZs are more easily activated in the interface than in
  the bulk glass. Thus, a closer look at the interface is warranted:
    We performed a Voronoi analysis of the atoms in the interface and
  glass phase. The interface is defined such that it includes glass atoms at the direct glass-crystal interface and their neighbors in the glass
  phase within a cutoff of $3.6 \, \rm${\AA}. 
  Figure~\ref{fig:snapshots}(f) exemplarily shows the Voronoi statistics of the Cu atoms in the glass and interface phase of the sample with network microstructure $\phi=170\, \rm \mu m$, both before and after 40ns of tensile creep testing. Prior to the Voronoi tesselation, an energy minimization using molecular statics has been performed. Here, only polyhedra that either occur more often than $2\%$ in the glass phase or interface phase are shown, the remaining polyhedra are summarized under the category ``others''.
When comparing the most prominent Cu centered polyhedra in the interface and glass phases, it is evident that the geometrically favorable motifs, 
also called Z-clusters~\cite{Ma2015}, are abundant in the glass. This is especially visible for the copper-centered icosahedra (Z12), and their defective version $\langle 0,2,8,2\rangle$.  Their fraction is significantly reduced 
in the interface, where more geometrically unfavored motifs (GUMs) \cite{Ding2014}, such as the $\langle 0,3,6,4 \rangle$ polyhedron, can be found. Additionally, 13-fold $\left <0,5,2,6 \right> $ 
and 14-fold  $\left < 0,4,4,6\right>$ clusters appear  at the immediate glass-crystal interface. The zirconium-centered polyhedra exhibit similar trends and are omitted here. In total, these results indicate a weakened 
glass structure in the interface. This is also supported by the lower number density of the interface of $50.1 \, \rm atoms/nm^{3}$ as
compared to $63.3 \, \rm atoms/nm^{3}$ in the glass matrix. Thus there, is excess volume in the interface.
Additionally, Fig.~\ref{fig:snapshots}(f) shows that the Voronoi statistics of the glass do not change during creep. In the interface, there is a slight increase
in occurence of Z12 and Z13 clusters after $40 \rm \, ns$ creep testing. Moreover, in a common neighbor analysis, we could observe
a minor increase of $0.4 \, \%$ in the fraction of bcc atoms in the overall composite. These last two observations indicate a marginal stress-assisted relaxation of the interface. 
Nonetheless, the effect is negligible.

  When comparing the composites in Figs.~\ref{fig:snapshots}(b)-(e), it can be seen that the higher $\phi$,
  the higher is the amount of glass atoms with a high atomic strain. This matches the observation of increasing creep rates with
  increasing $\phi$. The more STZ sites can be activated, the higher is the resulting creep rate.
    Nanoscale microstructures in Cu-Zr-based MGs have been reported in literature \cite{Hajlaoui2007, Pauly2010a, Pauly2010, Wang2014a}, with 
    secondary phase particle diameters between $2 \, \rm nm$ and $20\,\rm nm$. This corresponds to our spherical inclusions, which have a diameter of $9\,\rm nm$.
   In few studies with spherical precipitates, both the phase fraction and average diameter of the secondary phase are given \cite{Pauly2010a, Brink2016},
   which allows direct comparison of the specific interface area $\phi$ with our models.
   In Ref.~\cite{Pauly2010a}, which is closest to our geometry, the average diameter of these particles is around $20 \, \rm nm$ with
  volume fractions between $f=0.064$ to $0.0176$. This yields values of $\phi_{\rm exp}$ between $21/\rm \mu m$ to $53/\rm \mu m$.
  Since we use slightly higher volume fractions, we simulate higher specific interface areas, but
  they only differ by a factor of 4 to 10 from the experimental ones. Hence, we expect that the effect of increasing creep rate with increasing interface
is experimentally observable.
    
  \subsection{Influence of interface geometry}

\begin{figure}[h!]
      \centering   
    \includegraphics[width=8.6cm]{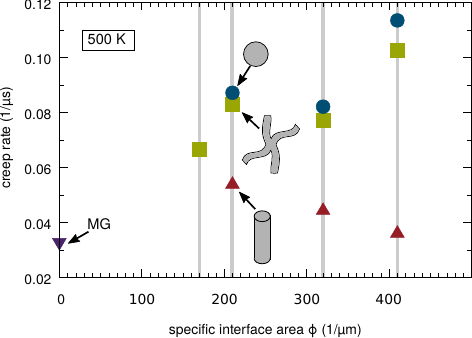}
    \caption{Correlation of creep strain rates fitted from creep stage II and specific interface area of various
    composite microstructures. }
\label{fig:strainrate}
\end{figure}

We showed above that the interface promotes
  creep and that the specific interface area $\phi$ is directly
  related to the creep rate. If we assume a simple, three-phase
  composite model of glass, crystal, and interface, this would suggest that
   $\phi$ is the only parameter for creep. We tested this by
  generating simpler geometric shapes with similar parameters $\phi$
  and $f$ to the network structures, where $f$ is always $0.3$. 
Considering that the network composites have a complex interphase topology with a variety of surface orientations, we at first construct two simpler geometries: 
A non-percolating crystalline phase in the form of spherical particles and a percolating phase made of nanowires aligned with the loading direction, 
(see Figs.\ \ref{fig:f1}(c) and~\ref{fig:f1}(f)). 
 These samples are then subjected to the same
load of $750 \, \rm MPa$ at $500 \, \rm K$ for $40 \, \rm ns$. Their
creep rates are shown in
Fig.~\ref{fig:strainrate}. If $\phi$ were the only parameter influencing the creep rates of our composites, the latter should 
remain unchanged when the shape of the crystalline phase is varied for constant $f$ and $\phi$. However, from Fig.~\ref{fig:strainrate} we see that this is not the case.  
Although all composite structures exhibit higher creep rates than the pristine glass, different secondary phase shapes lead to different creep rates for the same  $f$ and $\phi$.  
The following trend can be identified for the network and sphere-containing composites: The higher the specific interface area, the higher is the creep rate.
However, the opposite trend is visible for the nanowire composites.
Thus, a simple model which only takes into account the relative fraction of the different phases is insufficient.  Even the percolation cannot explain the different rates, as the non-percolating case (spheres) and the percolating network structures behave similarly.  The remaining difference between the three cases is therefore the interface orientation.

Thus, in the next step we studied the creep rate dependence on the interface geometry for constant $f=0.3$ and $\phi=210 \, \rm \mu m^{-1}$.
We also compared composites with percolating and non-percolating secondary phases: 
In addition to the secondary phase network, nanowires aligned with the loading direction and spherical particles,  see Figs.~\ref{fig:f1}(b), \ref{fig:f1}(c) and~\ref{fig:f1}(f), 
we introduce  non percolating nanorods as  in Fig.~\ref{fig:f1}(e) and  percolating nanowires, that are
  tilted $45^{\circ}$, see Fig.~\ref{fig:f1}(d), and $90^{\circ}$ to the $z$-axis. 
\begin{figure}[h!]
  \centering
  \includegraphics[width=8.6cm]{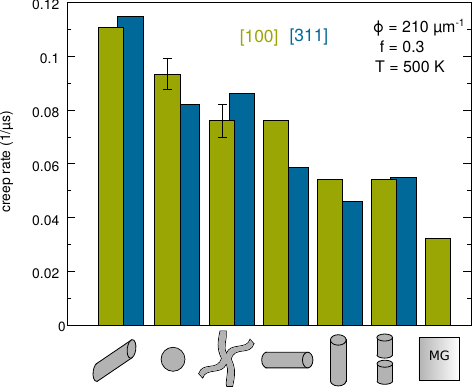}
  \caption{Creep rates of composites with $\phi_2 = 210 \, \rm \mu m^{-1}$ and volume fraction of $30\, \%$, but different shape of the secondary phase fitted from creep stage~II. 
  The different colors reflect the orientation of the crystalline phase with respect to the loading direction: Green means [100],
  blue means [311]. 
    }
   \label{fig:rates}
\end{figure}
In Fig.~\ref{fig:rates}, all these composites with different specific interface area $\phi$ are ordered with respect to their creep rates. 
The reference creep rate of the glass structure is shown on the right. 
The green bars are creep rates fitted from composites where the CuZr B2 crystal orientation is $[100]$. 
The highest creep rate is shown by the composite containing nanowires oriented $45^{\, \circ}$  with respect to the 
loading direction. This is followed by the sphere and network composite, that have a similar creep rate.
After that, the creep rate decreases for the wire that lies perpendicular to the loading direction. 
Furthermore, the lowest composite creep-rate is shown by  the wires oriented along the loading direction. 
Interestingly, they are not significantly changed when the wires are non-continuous. The error bars for spheres represent the 
standard deviation of four different spatial distributions of the spheres, while the error bars for the networks were determined by 
loading the structure in the three cartesian directions.  We note that the size of the error bars is comparable to the difference in 
rates between the nanowires in Fig.~\ref{fig:strainrate}.  The reduction of rate with interface area does therefore not exceed the
margin of error.  The fluctuations in the creep rate can be explained by the fact that the glass is heterogeneous on length scales
of nanometers~\cite{Ma2015} and is therefore sensitive to the location of the crystalline phase.  Nevertheless, the hierarchy 
presented in Fig.~\ref{fig:rates} exceeds the margin of error.

As the interface controls the mechanical behavior, the results of our simulations may be influenced by the orientation of the crystals.
Hence, we repeated the simulations with B2 crystalline phases oriented with the $[311]$ direction along the $z$--axis. The results are 
also shown in Fig.~\ref{fig:rates} and do not significantly deviate from the results with a $[100]$ orientation.  Indeed, it was found
earlier that the interface energy of copper crystals in $\mathrm{Cu}_{64}\mathrm{Zr}_{36}$ is only weakly dependent on the crystal 
orientation~\cite{Brink2015}, which is in line with the current results.

\section{Interface Schmid factor model}
\label{sec:Schmid}
\begin{figure}[h!]
    \centering   
    \includegraphics[width=7cm]{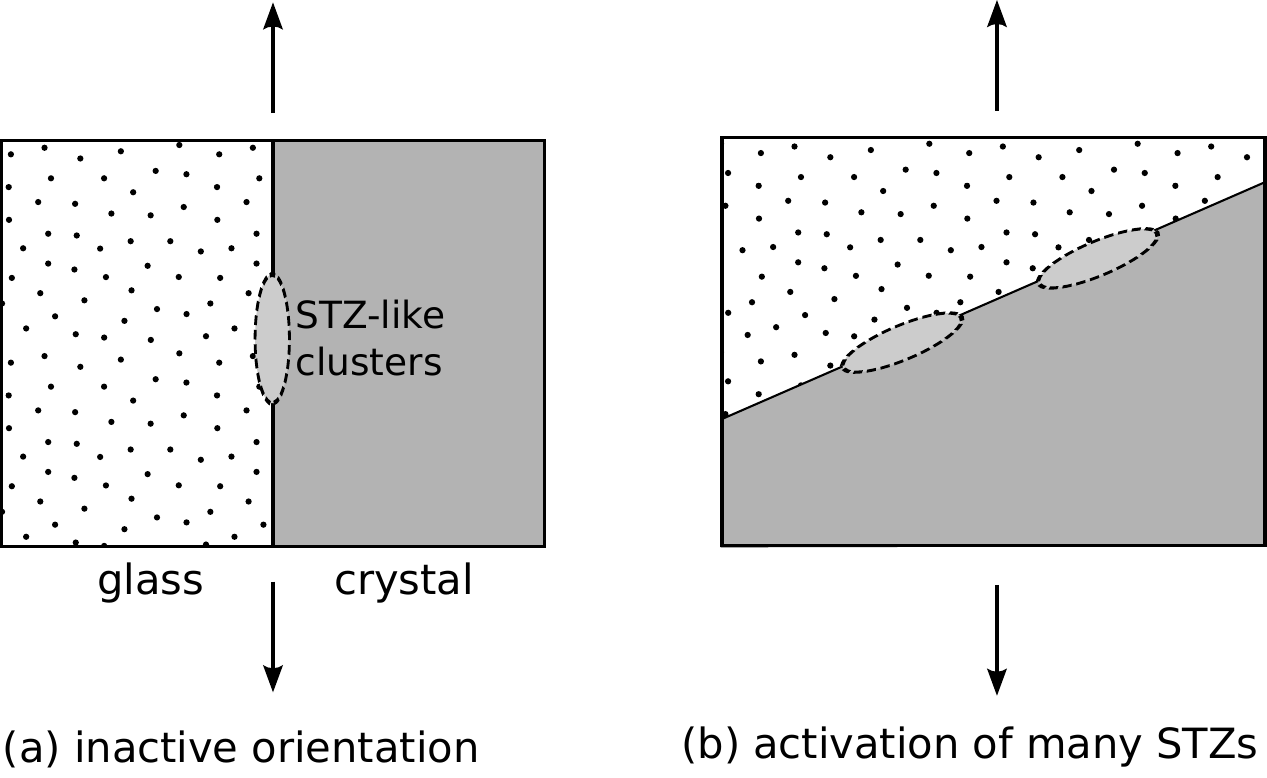}   
    \caption{The more interfaces with
    the highest Schmid factor are present in the composite, i.e.,\ interfaces oriented  $45^{\, \circ}$  to the tensile load, the more
    STZs are activated in the glass. }
\label{fig:stz}
\end{figure}

The alignment of the glass--crystal interfaces is therefore the main factor controlling the creep rate, while percolation only seems 
to play a minor role. For the nanowires oriented in loading direction, the interface normal is always perpendicular to the loading 
direction, see Fig.~\ref{fig:stz}a. The $45^{\circ}$ orientation however, seems to be favorable for creep deformation (Fig.~\ref{fig:stz}b).
Based on the idea of the Schmid factor  $m$ in a crystal, which translates the applied stress $\sigma$ into a resolved shear stress $\tau$ on
 the different slip planes and glide directions in a crystal, we define a Schmid-like factor for our composites. In a crystalline material, it is
 \begin{eqnarray}
\tau& =&m \cdot \sigma  \\
&= & \cos\kappa \cdot \cos \lambda \cdot \sigma \; .
\end{eqnarray}
In our case, the amorphous interface regions act as preferential slip sites, so we use the concept of the Schmid-factor in a similar geometric derivation.
\begin{figure}[t]
  \centering
  \includegraphics[width=8.6cm]{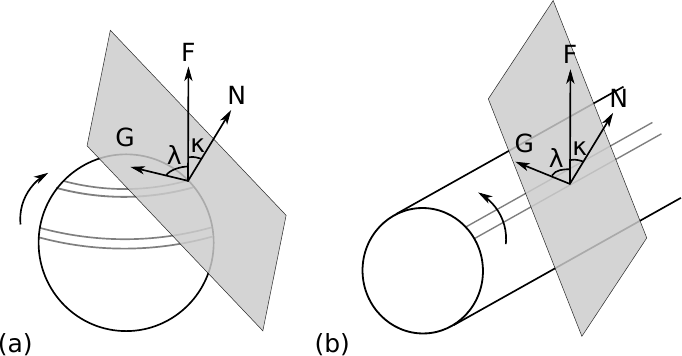}	
  \caption{ The integrated ``interface'' Schmid factor is expected to be different on a sphere surface (a) and cylinder (b). $F$ is the direction of 
  load, $N$ is the normal of the tangential plane and $G$ is the 
  glide direction. }
\label{fig:schmid-calc}
  \end{figure}
However, we assume that in the glassy interphase there are no restrictions to the glide direction $\vec G$ on the slip plane. The glide direction is thus the direction that
maximizes $\cos\lambda$, which corresponds to the projection of the externally applied force $\vec F$ on the slip plane (represented by its normal vector $\vec N$).
Using the definition of $\cos \kappa$ from the classical Schmid law, we arrive at
\begin{eqnarray}
m(\vec{N}, \vec{F}) &=& \cos\kappa \cdot \cos \lambda \\
  \label{eq:glide-dir1}
& =& \frac{\vec N \cdot \vec F }{ |\vec{N}| |\vec{F}|} \cdot 
\underbrace{\frac{ \vec{G} \cdot \vec{F}}{ |\vec{G}| |\vec{F}|}}_{\rm max} \; ,  \\
  \label{eq:glide-dir}
  \text{with } \vec G &=& \vec F - \frac{\vec F \cdot \vec N}{ | \vec N |^2  } \vec N \; .
\end{eqnarray}
In Fig.~\ref{fig:schmid-calc} two example interface geometries are sketched.

\begin{figure}[h!]
  \centering
  \includegraphics[width=8.6cm]{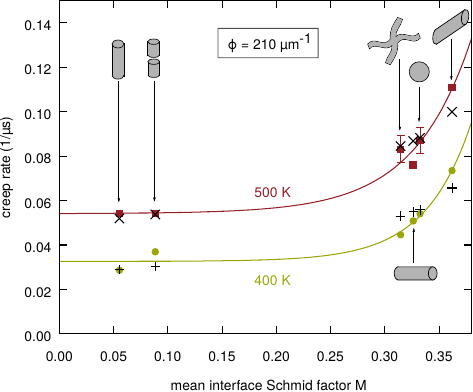}	
  \caption{ The data points show the creep rates vs. mean interface Schmid factor of composites with different shapes of the secondary phase but constant $\phi$ and $f=0.3$. As a guide to the eye, 
  the lines show an Argon-like creep behavior where the resolved shear stress is substituted by $M\cdot \sigma$. In addition, there is an offset caused by the fact that the glass matrix contributes to the creep strain. To properly fit the activation energy for creep, however, the local Schmid factors $m_i$ have to be used, 
  as they are included in the $\sinh$ function, see Eq.~\eqref{eq:argonfit}. The results are plotted as black crosses.
}
\label{fig:schmid}
\end{figure}

We expect that different secondary phase topologies have different distributions of these local Schmid factors.
In order to characterize these topologies, we use a total mean Schmid factor
\begin{equation}
  \label{eq:average-m}
M = \frac{1}{ A} \int_A m (\vec N_{A'}, \vec F) \, \mathrm{d} A' \; .
\end{equation}
(Note that $m$ depends on the normal vector of the specific differential area for which it is calculated and cannot be moved out of the integral.)
We choose six different composite structures with $\phi=210 \, \rm \mu m^{-1}$ and $f=0.3$ for comparison. The integration over
the interfaces is performed numerically: The software \textsc{ovito}  is used to create a surface mesh of the interface region by Delaunay triangulation~\cite{Stuko2014}. For each of these triangles, 
we can calculate the surface area $A$ and the surface normal $\vec N$ and thus $m$. Figure~\ref{fig:schmid} shows the resulting total interface Schmid factors on the abscissa,
while the ordinate shows the corresponding creep rates for the samples with $\phi=210 \, \rm \mu m^{-1}$.
The data points show that---at constant temperature---the composite structures with higher total interface Schmid factors $M$ exhibit higher creep rates. 
Note that since the mean interface Schmid factor has been determined numerically, even for the nano-rods and nanowire-type inclusions, $M$ is not equal to zero. This can be attributed to the fact that 
there are surface triangles with surface normals non-perpendicular to the loading direction at the edges.
The lines in Fig.~\ref{fig:schmid} result from a fit of the Argon model [Eq.~\eqref{eq:argon}] using $\tau = M\sigma$. This is obviously only a very rough trend, since the resolved
stress varies for each surface triangle, each of which therefore contributes differently to the total creep. As such, we employed a model using local resolved stresses to obtain quantitative results:
\begin{equation}
\begin{split}
\label{eq:argonfit}
\dot \epsilon (T,m_1,m_2,\ldots) & =  \underbrace{\alpha \nu_0 \gamma_0}_{\rm const.}  \cdot \exp \left( -\frac{Q}{kT} \right) \cdot   \\ 
 & \cdot   \left < \sinh \left(  \underbrace{ \gamma_0 \Omega}_{\rm const.}  \cdot \frac{ m_i \sigma }{ kT} \right) \right >_{\!\!i}
 +  \dot \epsilon_{0}(T) \; .
 \end{split}
\end{equation}
In this formulation, the shear stress $\tau$ is replaced by the local resolved shear stress $m_i \sigma$, calculated for each surface triangle $i$. This modified $\sinh$ term is then averaged.
This part of the equation represents the behavior of the interface phase. The term $\dot \epsilon_{0}(T)$ includes the contributions of the glass phase, while the crystalline phase only responds
elastically and is therefore not included in the formula. Fitting parameters are the prefactor $\alpha\nu_0\gamma_0$, the activation barrier $Q$, the STZ activation volume $\gamma_0\Omega$,
and $\dot\epsilon_0(T)$. The results of the fit, including simulations at two different temperatures, are plotted using crosses in Fig.~\ref{fig:schmid} and agree well with the data. Table~\ref{tab:I} lists the results of the temperature independent fit parameters.

As we can see, the activation barrier in the interphase is severely reduced compared to values found in literature for the homogeneous glass \cite{Schuh2007, Argon1980}. 
However, we observe that the noise in the creep curves is quite large compared to the creep rates that we try to fit. A detailed treatment of this problem can be found in the 
appendix. The bottom line is that we cannot draw a quantitative conclusion concerning the activation energy and the prefactors.
Nonetheless, this model shows that the interphase can be described as a weaker glass, where STZs are more easily activated. 
As such, it is more susceptible to a creep-like deformation behavior, even at lower temperatures. The orientation dependence hints towards anisotropic creep behavior of composites.
\begin{table}[t]
 \caption{Temperature independent fit parameters in Eq.~\eqref{eq:argonfit}.}
 \label{tab:I}
 \renewcommand{\arraystretch}{1.5}
 \begin{tabular}{|p{1.6cm} || p{2.0cm} | p{1.6cm}| p{1.8cm}|}
\hline
parameter & $\alpha \nu_0 \gamma_0$ (GHz) & $ Q $ (eV)& $\gamma_0 \Omega$ $\rm (nm^3)$\\ 
\hline
 fit 	& $0.394 $ & $0.346$ & $0.172 $  \\
  \hline 
\end{tabular} 
\end{table}

\section{Elastic and plastic creep deformation}
\label{sec:plasticity}
\begin{figure}[h!]
    \centering   
    \includegraphics[width=8.6cm]{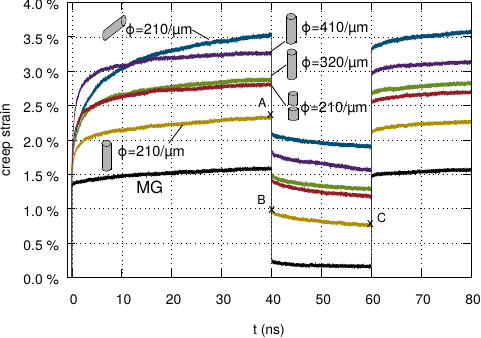}
    \caption{Loading--unloading cycle for nanowire composites.  The distance AB marks the elastic part of the creep deformation. BC is a recoverable anelastic part and C the remaining plastic deformation.}
\label{fig:loadingcycle}
\end{figure}

So far, we only quantified the composites' creep behavior in terms of creep rates.
In addition to the creep rates fitted from steady state behavior, 
it is important to understand what happens at the onset of creep.
Usually, the individual elastic and plastic contributions of the creep
deformation overlap and are difficult to distinguish in the
creep curve. The remaining plastic deformation can, however, easily be
identified from the deformation that remains after unloading. Therefore, we revisit the creep simulations of the nanowire composites with $\phi_2$, $\phi_3$ and $\phi_4$. After $40 \, \rm ns$, the
wire composites are unloaded and relaxed for $20 \, \rm ns$. The corresponding data are depicted in
Fig.~\ref{fig:loadingcycle}.

  \begin{figure}[h!]
    \centering   
    \includegraphics[width=8.6cm]{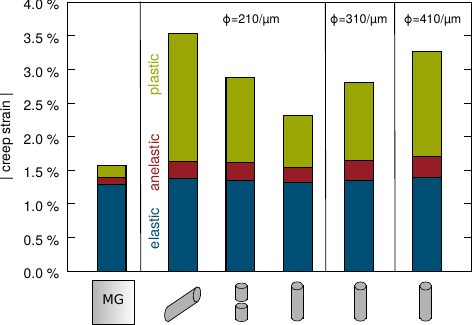}
    \caption{ Different contributions to the strain calculated from the strain drop after unloading in Fig.~\ref{fig:loadingcycle}.}
\label{fig:straincomponents}
\end{figure}
  
 The difference in strain between points A
and C marked in Fig.~\ref{fig:loadingcycle} shows the composites'
reversible part of the creep deformation. While AB is the elastic
strain instantly recovered when the load is removed, BC shows an
anelastic part that can be recovered over time. What is striking is that for all composites
a significant part of the deformation is reversible; the homogeneous glass exhibits a lower anelastic strain. For the crystal, we would expect that once the load is removed it contracts to its initial length. In the composite, however, this is 
hindered by the glass-phase. We conclude that this remaining back stress together with a possible higher anelasticity of the interface leads to a significant time-dependent strain recovery. 
We note that the anelastic recovery is not completed after $20 \, \rm ns$, but that the anelastic recovery after $20 \, \rm ns$ is very similar for all composites, see Fig.~\ref{fig:straincomponents}. 
The same is true for the elastic part, which is also comparable to the homogeneous glass.
The plastic deformation that remains after removing the load and subsequent relaxation for $20\, \rm ns$, though, shows a significant dependence on the  shape of the crystalline phase. 
This is in agreement with the results of the previous sections, which are now shown to be clearly plastic in nature, as the anelastic and elastic contributions do not differ significantly between the samples.
This is to be expected, since the STZs (which are sensitive to the factor $M$) are plastic events.

\section{Conclusion}

In our simulations of the creep behavior of $\rm Cu_{64}Zr_{36}$-glass and B2-crystal composites, we could observe a mixed regime of elastic, anelastic and plastic deformation.
When comparing the creep behavior of the homogeneous glass to glass--crystal composites, where the crystalline phase only responds elastically, it could be shown that
the composites have much higher creep rates than the homogeneous glass. Clearly, creep is promoted by the presence of interfaces. In our model system the interface presents a weak spot in comparison to
the surrounding MG. Furthermore, by keeping the volume fraction constant but varying the shape of the secondary phase, we could analyze the influence of the specific interface area and the percolation of
the secondary phase. We found that the creep rates of the composites strongly depend on the orientation of the glass--crystal interface. We have given an explanation in terms of the resolved shear stress $\tau$
derived from the concept of the Schmid factor in crystalline materials. This dependence  of the creep rate on the anisotropy of the secondary phase should be taken into account when  designing
creep resistant MG composites.

\section*{Acknowledgments}
The authors gratefully acknowledge the financial support of the
  Deutsche Forschungsgemeinschaft (DFG)\linebreak through project grants STU~611/1-1 and RO~4542/2-1. 
  Computing time was provided by Technische Universit\"at
  Darmstadt on the Lichtenberg High Performance Computer.

\appendix
\renewcommand*{\thesection}{\appendixname~\Alph{section}} %

\section{Quantitative reliability of fits to\\ simulation data}

\begin{figure}[h]
    \centering   
    \includegraphics[width=8.6cm]{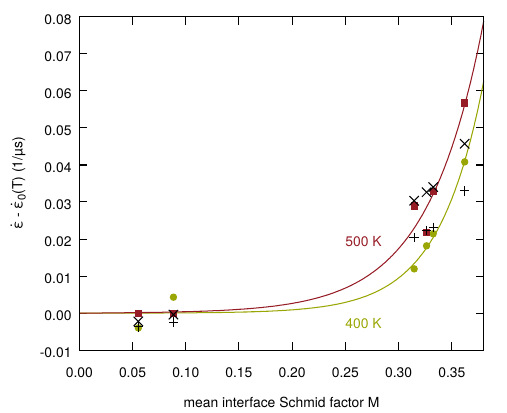}
    \caption{Colored data shows the creep rates and mean interface Schmid factors treated in section~\ref{sec:Schmid}, but without the offset term. Thus, they reflect solely the properties of the interface. Black crosses represent the fit to the creep rates using Eq.~\eqref{eq:argonfit}.}
\label{fig:A2}
\end{figure}

The activation barrier $Q$ can only be determined by including data
obtained using different temperatures and crystalline phase
geometries, characterized by $M$. If we remove the temperature
dependent offset term $\dot \epsilon_0(T)$ in the composites' creep
rates, we note that the data points at the two temperatures are very
close together, see Fig.~\ref{fig:A2}. The scatter is on the order of
the difference between the two data sets. Thus, the activation energy
and the prefactors of the interface phase can only be determined
approximately and cannot be interpreted as quantitatively reliable.

\end{document}